\documentclass[aps,prl,twocolumn,showpacs,floatfix]{revtex4}
\usepackage{amsmath}
\usepackage{graphicx,epsfig,psfrag,color,colordvi}
\usepackage{amssymb}

\begin{document}
\title{Quantum phase transition in the two-band Hubbard model}
\author{T. A. Costi and A. Liebsch}
\affiliation{Institut f\"ur Festk\"orperforschung, Forschungszentrum J\"ulich,
52425 J\"ulich, Germany.}

\date{\today}
\begin{abstract}
  The interaction between itinerant and Mott localized electronic states in strongly 
  correlated materials is studied within dynamical mean field theory in 
  combination with the numerical renormalization group method. 
  A novel nonmagnetic zero temperature quantum phase transition is found in the 
  bad-metallic orbital-selective Mott phase of the two-band Hubbard model, 
  for values of the Hund's exchange which are relevant to typical transition 
  metal oxides. 
\end{abstract}
\maketitle

The nature of the metal insulator transition in multi-component systems, with 
competing energy scales associated with differing bandwidths and Coulomb and
exchange interactions, is currently of great interest~\cite{imada.98}. For instance, in 
transition metal oxides such as Ca$_{2-x}$Sr$_{x}$RuO$_{4}$
or manganites (e.g. La$_{1-x}$Sr$_{x}$MnO$_3$), 
the Coulomb interaction between weakly and strongly correlated subbands can 
give rise to complex phase changes as a function of temperature, pressure,
or impurity concentration. 
An interesting aspect of such systems is the coexistence of itinerant wide
band electrons with partially or completely localized narrow band electrons. 
This gives rise to bad-metallic 
behavior as observed, for example, in the resistivity of
the paramagnetic phases of VO$_{2}$~\cite{allen.93} and 
Ca$_{2-x}$Sr$_{x}$RuO$_{4}$ at $x=0.2$~\cite{nakatsuji.00}. 
Even in standard Mott insulators such as V$_2$O$_3$
and LaTiO$_3$, the presence of inequivalent orbitals leads to strong orbital
polarization where in the vicinity of the transition nearly localized electrons 
coexist with weakly itinerant bands~\cite{keller.04,poteryaev.07,pavarini.04}. 
In the case of cuprates, 
one of the most fascinating aspects is the coexistence of 
strongly and weakly correlated
regions of the Brillouin zone, giving rise to so-called hot spots and cold spots,
with breakdown of Fermi-liquid behavior in the former and Fermi-liquid
properties in the latter~\cite{civelli.05,kyung.06,zhang.07}. 
Since the momentum variation of the self-energy is 
associated with spatial fluctuations, an intriguing analogy exists between
multi-site interactions within a single band and single-site interactions 
among inequivalent orbitals.

A simple model which captures some of the essential physics occurring in the systems 
mentioned above is the two-band Hubbard model consisting of narrow and wide subbands, 
coupled via local Coulomb energy $U$ and Hund's exchange 
$J$~\cite{anisimov.02,liebsch.04}.     
As a result of the various energy scales contained in this model, its phase diagram 
turns out to be remarkably rich, as shown in detailed studies by many authors~%
\cite{liebsch.05,koga.04,ferrero.05,de-medici.05,ruegg.05,arita.05,%
inaba.05,knecht.05,biermann.05,inaba.06}. These studies revealed that 
differing bandwidths and a finite Hund's exchange stabilize an orbital-selective
Mott phase, in which the wide band itinerant electrons coexist with 
localized spins arising from Mott localization of the narrow band electrons. 
Further increase of the Coulomb energy $U$ leads to a Mott transition in the
wide band whose nature has remained unresolved for the important case of
anisotropic Hund's exchange of relevance to many transition metal compounds.

The aim of the present work is to identify the character of the wide band 
transition of the two-band Hubbard model at zero temperature for anisotropic
Hund's exchange. We use DMFT~\cite{dmft} in combination with the numerical 
renormalization group (NRG) method~\cite{nrg,costi,bulla.99,hofstetter.00}  
in order to achieve a sufficiently accurate description of low-frequency properties.
We find that anisotropic Hund's exchange yields 
a continuous quantum phase transition as long as $J/U > j_{c}$, where 
$j_{c}\approx 0.1$. This range of $J/U$ values is typical of transition
metal oxides (e.g., for manganites). In this region, the spectral density 
$A(\omega)$ vanishes at the Fermi level as $|\omega|^{\delta}$, 
with $\delta=1/3$. This behavior
coincides with the one for the metal insulator transition in the exactly solvable 
limit $U=0$, with variable $J$. In striking contrast, the region $0<J/U < j_{c}$ 
exhibits a first-order Mott transition, which differs from the continuous 
zero temperature Mott transition for the one-band model~\cite{dmft}.

We consider an effective low-energy model specific for the 
orbital-selective Mott phase~\cite{biermann.05}  which allows 
studying the transition of the wide band in detail. 
In this model, the electrons in the
narrow band of the original two-band Hubbard model have undergone 
Mott localization so their low-energy degrees of freedom are represented 
by localized spins.  These couple to the itinerant wide 
band electrons via a Hund's exchange. The resulting model is the 
ferromagnetic Kondo lattice model with interactions in the band,
\begin{eqnarray}
\nonumber
H_{\rm fkl}&=&-\sum_{ij\sigma} t_{ij} c^\dagger_{i\sigma}c_{j \sigma}
+ U\sum_{i} n_{i\uparrow}n_{i\downarrow}\\
&-\,&\sum_{i} \left[2J\;S^z_{i}s^z_{i}+2J'(S^{+}_{i}s^{-}_{i}+S^{-}_{i}s^{+}_{i})\right]
\label{eq:fkl}
\end{eqnarray}
where $s^{z}_{i}$ and $s^{\pm}_{i}$ are the z-component and raising/lowering 
operators, respectively, for electrons in the wide band at site $i$ and 
$S^{z}_{i}$ and $S^{\pm}_{i}$ are the corresponding
operators for the localized spin at site $i$. The anisotropy of the Hund's exchange is
measured by $J'/J\le 1$ with $J'=J$ corresponding to isotropic
and $J' = 0$ to Ising Hund's exchange.  The effective impurity model corresponding
to (\ref{eq:fkl}) which we solve within DMFT-NRG is the ferromagnetic 
Kondo model with a local Coulomb repulsion.
As in previous works we used a Bethe-lattice density of states with half-bandwidth 
$D=1$, which we henceforth take as our unit of energy. In the NRG
procedure $800-1200$ states per energy shell were retained. The 
logarithmic discretization parameter for the conduction band was 
$\Lambda=1.5$ and $N=90$ shells were used, corresponding to a 
minimum energy resolution of $\Lambda^{-(N-1)/2}\approx 10^{-8}$. 
The calculations have been carried out for fixed values of $J/U$ and increasing values
of $U/D$, as we consider the situation where correlations are tuned by
reducing the bandwidth $D$ whereas the ratio $J/U$ is material dependent, being,
for example, larger than unity for manganites and smaller than unity for
most transition metal oxides. As we are interested in the metal insulator 
transition we consider a half-filled band and paramagnetic solutions only.

Note that the model (\ref{eq:fkl}) can also be viewed as
a single-band Hubbard model in the presence of spin-disorder scattering of
strength given by the Hund's exchange. 
For $U=0$ and $J>0$ the model reduces to the double exchange
model used to describe manganites which is known to exhibit non-Fermi liquid 
behavior on restricting to paramagnetic groundstates~\cite{furukawa.94}. 
This non-Fermi liquid or bad-metallic 
behavior persists to $U>0$~\cite{liebsch.04,biermann.05,liebsch.06}.

\begin{figure}
\includegraphics[width=0.30\textwidth,angle=-90]{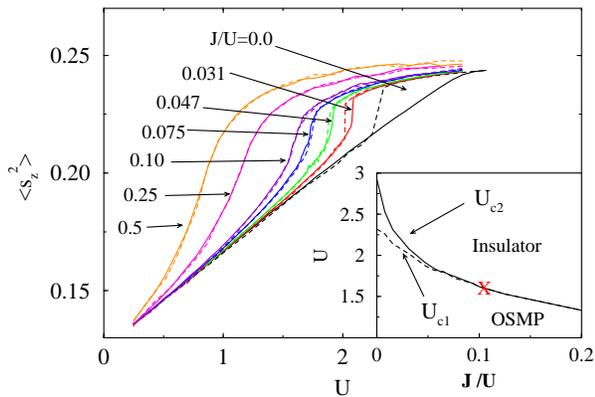}
\caption
{(Color online) Effective magnetic moment $\langle s_{z}^{2}\rangle$ 
per site of the conduction electrons for several values of $J/U$, $J'=0$ 
as a function of $U$. Solid (dashed) lines: increasing (decreasing) 
values of $U$.
For $0<J/U<0.1$ the Mott transition is of first order with
hysteresis in a region $U_{c1}<U<U_{c2}$ (the lines are guides to the eye,
a finer grid of $U$ points, as used for $J/U=0.031$, 
reveals vertical jumps at $U_{c1,2}$). For $J/U>0.1$ the transition becomes
a continuous quantum phase transition. Inset: $T=0$ phase
diagram and dependence of $U_{c1,2}$ on $J/U$. OSMP denotes
the orbital-selective Mott phase.  \Red{X} denotes the critical 
point $J/U=j_{c}\approx 0.1$, $U=u_{c}\approx 1.6$
where the Mott transition changes from first to second order.
}
\label{fig1}
\end{figure}

In the present work we focus on the Ising limit, i.e., $J'=0$.
While this might appear as a severe restriction, this limit is actually 
relevant for all cases with $J'<J$, and therefore should be important in 
materials where as a result of non-cubic octahedral distortions Hund's 
exchange will not be fully isotropic. This can be understood from 
the well known renormalization group flow of the ferromagnetic 
Kondo impurity model~\cite{anderson.70} and explicit numerical calculations
including the local Coulomb interaction $U$~\cite{costi.07}. 
For a metallic band, these predict that for anisotropic Hund's 
exchange, $J' < J$, the spin-flip part of the exchange, $J'$, 
renormalizes to zero, but the Ising part, $J$, renormalizes to 
a finite value $J^{*}>0$. Similarly, for an insulating band, 
one finds that the Ising coupling renormalizes to a finite 
value and the system is gapped so $J'<J$ becomes irrelevant 
for low energy properties. From these arguments it follows that, 
for the purposes of determining the nature of the Mott transition 
in the wide band for anisotropic Hund's exchange, it suffices to 
consider just Ising anisotropy $J'=0$. Results for general $J'$ will be given 
elsewhere~\cite{costi.07}. A finite Ising exchange for
metallic and insulating cases also implies that the entropy of the
lattice model (\ref{eq:fkl}) is $\rm{ln 2}$  per site for both metallic
and insulating states. Consequently, the metal insulator transition
at finite temperature need not be first order, and in particular
a continuous quantum phase transition can arise at $T=0$ which becomes
a crossover at finite temperature. We also note that the 
renormalization of $J'$ to $J^{*}=0$ in the metallic case implies
that the spin-disorder in the lattice model (\ref{eq:fkl}) 
persists to $T=0$, giving rise to above mentioned bad-metal behavior with 
$\rm{Im}[\Sigma(\omega=0)] < 0$~\cite{liebsch.04,biermann.05,liebsch.06}, 
where $\Sigma$ is the wide band conduction electron self-energy.

To provide an overview of the different transition regimes of the model 
defined in (\ref{eq:fkl}) we show first in Fig.~\ref{fig1} the behavior 
of the effective conduction electron magnetic moment, $\langle s_{z}^{2}\rangle$, 
as a function of increasing/decreasing $U$ for several fixed values of
$J/U$. For $J/U=0$ we recover the Mott transition in the single-band Hubbard
model with a hysteresis region and a continuous metal insulator transition
at $U=U_{c2}\approx 3$. The hysteresis behavior persists for small but 
finite exchange although it rapidly diminishes with increasing $J/U$ 
(see inset to Fig.~\ref{fig1}). 
Within the numerical accuracy of our calculations, the hysteresis 
behavior vanishes for $J/U > j_{c}\approx 0.1$, indicating that the
Mott transition in this region is a continuous quantum phase transition. 
The critical Coulomb interaction for the Mott transition is
seen to decrease monotonically with increasing $J/U$ 
(inset to Fig.~\ref{fig1}). 
The precise location and nature of the critical 
point at $J/U=j_{c}\approx 0.1$, $U=u_{c} \approx 1.6$ where first and
second order lines meet requires a detailed study and is outside the scope of the
present paper. The phase diagram shown in Fig.~\ref{fig1} is consistent
with ED-DMFT results at low $T$ which  also yield first-order behavior for
$J/U<0.1$, $U<u_{c}\approx 1.6$ and a continuous transition 
for $J/U>0.1$, $U>u_{c}$~\cite{costi.07}.

\begin{figure}
\includegraphics[width=0.30\textwidth,angle=-90]{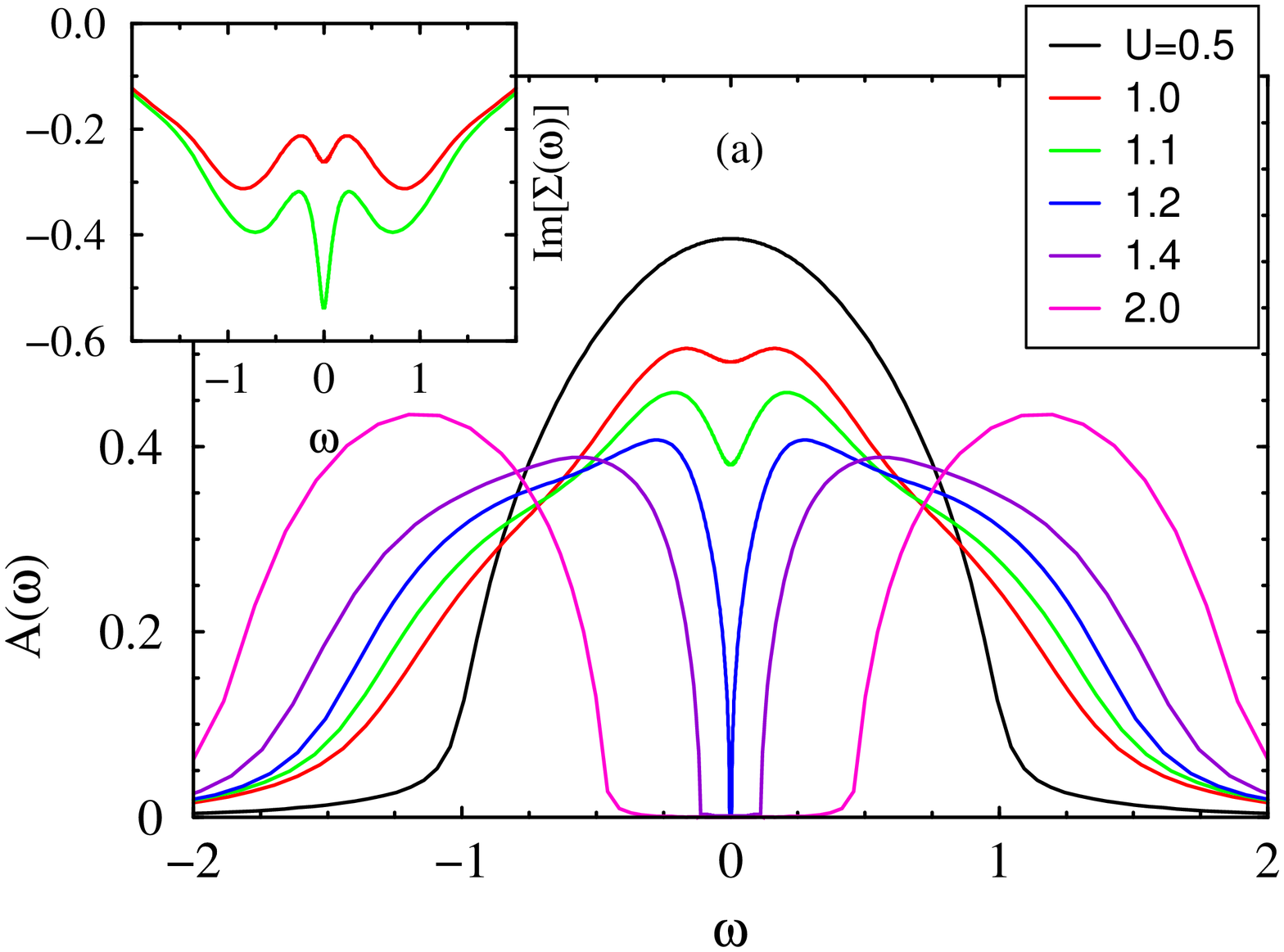}
\includegraphics[width=0.30\textwidth,angle=-90]{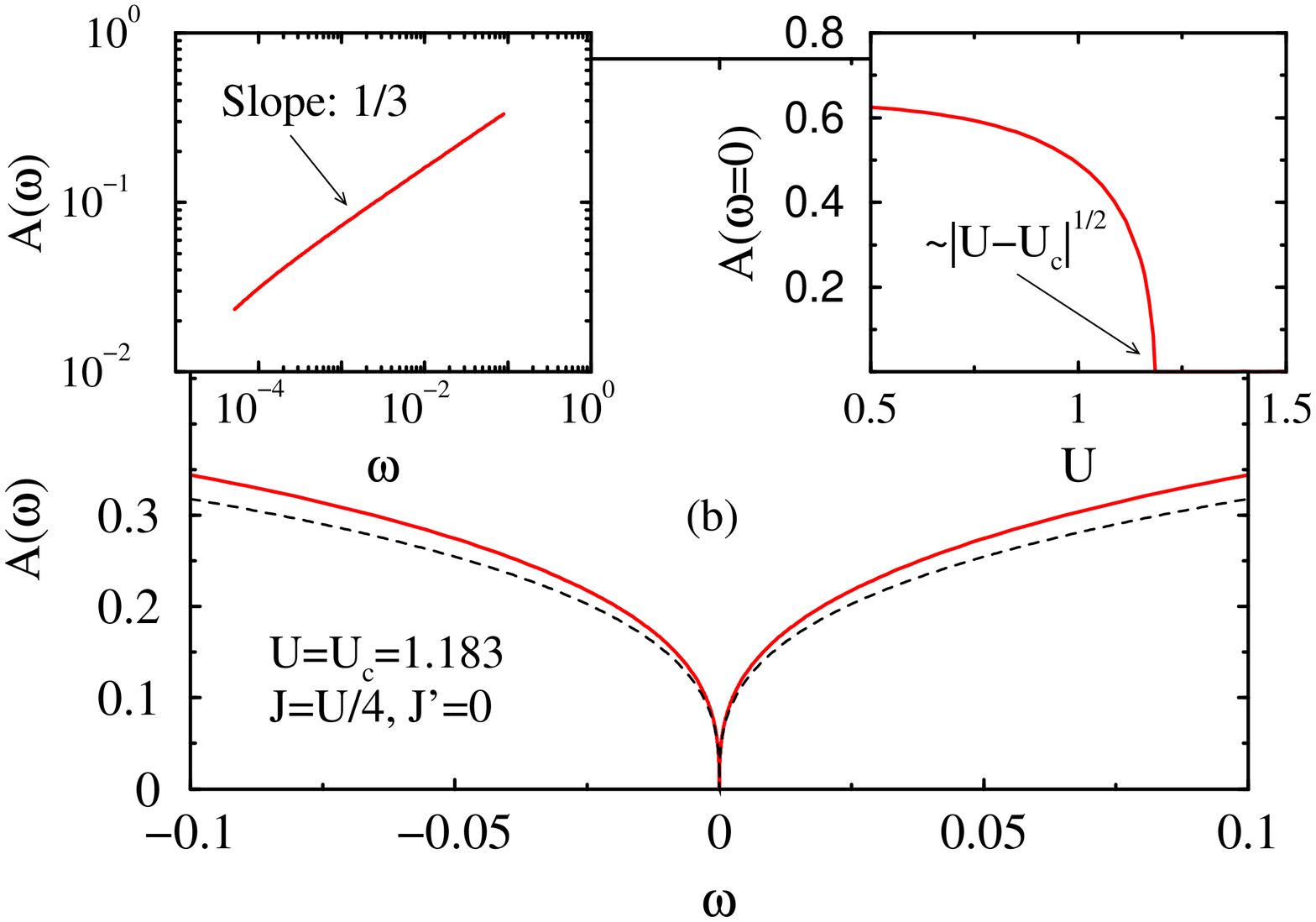}
\caption
{(Color online) Upper panel:
Spectral density of wide band for several values of $U$,
for $J=U/4$,  $J'=0$. The Mott transition occurs at 
$U_{c}=1.183$ and is continuous. Inset: imaginary part of
the self-energy for $U<U_{c}$, indicating a bad metal.
Lower panel: Low-frequency region of
spectral density $A(\omega)$ for $J=U/4$, $J'=0$ at 
$U_{c}=1.183$ (red solid line; logarithmic plot left inset); the
analogous spectrum for the exactly solvable model at $J_{c}=2$,
$U=0$, is shown by the dashed line. Right inset: $A(0)$ as a
function of $U$.
}
\label{fig2}
\end{figure}

Let us now look closer at the continuous quantum phase transition for $J/U > j_{c}$. 
Fig.~\ref{fig2}(a) shows the evolution of the spectral density as a function of $U$
for fixed $J/U=1/4$. For $J'\approx J$, 
this would be a value typical for transition metal oxides which 
has been used in many previous calculations~\cite{liebsch.04,liebsch.05,%
koga.04,ferrero.05,de-medici.05,ruegg.05,arita.05,knecht.05,inaba.06}. 
At small $U$, the density of states at the Fermi level, $E_{F}=0$, 
satisfies the pinning condition, $A(0)=2/\pi$, approximately. Near $U\approx 1$ 
the spectrum begins to exhibit a pseudogap which gets progressively deeper with 
increasing $U$. The characteristic scale for this pseudogap is $J$. 
At $U_{c}=1.183$ this pseudogap becomes a soft gap
before a full insulating gap opens at $U>U_{c}$.  This critical value
is in reasonable agreement with ED-DMFT results for $T=0$~\cite{koga.04} and 
$T>0$~\cite{liebsch.05}. In the bad-metallic region below the Mott transition, 
the electronic lifetime $\tau(\omega=0)\sim -1/{\rm Im} [\Sigma(\omega=0)]$ 
is finite (see inset to Fig.~\ref{fig2}(a)) and vanishes at $U=U_{c}$. 
Note that for this relatively large value of $J/U$, 
well defined upper and lower Hubbard bands are only formed once 
the insulating gap has opened. In contrast, for smaller values of $J/U$
well defined Hubbard bands are formed already in the metallic state, see 
Fig.~\ref{fig3} below. These $T=0$ spectra  are 
consistent with those derived within Quantum Monte Carlo (QMC) 
DMFT at finite $T$~\cite{liebsch.04,knecht.05,liebsch.06b}.
Because of the presence of the pseudogap the spectra exhibit a characteristic 
four-peaked structure, with low-energy features limiting the gap and high-energy
peaks associated with Hubbard bands.
 
We have analysed the low-frequency behavior of the spectral density
at the quantum phase transition in the region $J/U>j_{c}$ and 
find $A(\omega)\sim |\omega|^{\delta}$ with $\delta=1/3$, as
shown in Fig.~\ref{fig2}(b). Moreover, $A(0)\sim |U-U_{c}|^{1/2}$ for
$U\rightarrow U_{c}^{-}$. Remarkably, this behavior coincides with the
one obtained for the continuous Mott transition in the exactly solvable
model $J/U\rightarrow\infty$ with $U\rightarrow 0$ on increasing $J$.
Solving the DMFT equations for this model we find a band-splitting 
transition at $J_{c}=2$ with $A(\omega)\sim |\omega|^{1/3}$
and $A(0)\sim |J-J_{c}|^{1/2}$ for $J\rightarrow J_{c}^{-}$~\cite{vanDongen.97}.
This suggests that the low-frequency excitations in the whole region
$J/U > j_{c}$ are governed by the Ising Hund's exchange and that the 
paramagnetic metal insulator transition belongs to the same 
universality class as the band-splitting transition in the model 
with $U=0$ on increasing $J$.
\begin{figure}
\includegraphics[width=0.30\textwidth,angle=-90]{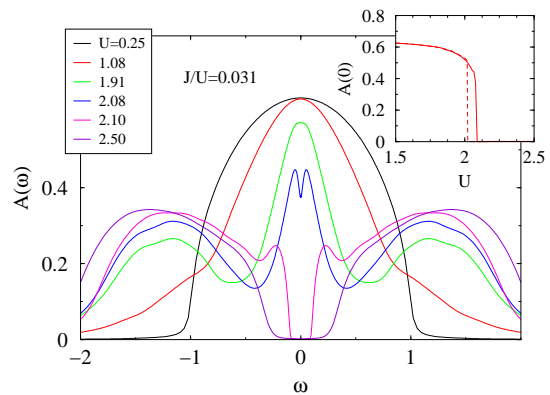}
\caption
{(Color online)
Spectral densities of the wide band for several values of $U$, for, 
$J/U=0.031$, $J'=0$. The Mott transition is first order and occurs at 
$U_{c}\approx 2.09$. Inset: $A(0)$ as a function of $U$ showing hysteresis
and discontinuous jump. Solid (dashed) lines: increasing (decreasing)
values of $U$.
}
\label{fig3}
\end{figure}
 
We now discuss the region $J/U < j_{c}$ where the effective two-band model
(\ref{eq:fkl}) yields a first-order Mott transition at $T=0$.
This is evidenced by the discontinuity in $\langle s_{z}^{2}\rangle$ 
at both $U_{c1}$ and $U_{c2}$ (see Fig.~\ref{fig1}) and 
in the discontinuous jump in the spectral
density at $E_{F}$, (see inset to Fig.~\ref{fig3}). 
A comparison of the groundstate energies of metallic and 
insulating solutions in the hysteresis regions shows that 
the metallic solution is lower in energy than the insulating
one, so the first-order transition at finite $J/U$ occurs at $U_{c2}$ as for
the $T=0$ metal insulator transition in the single-band model.
The spectral densities for $J/U=0.031$ shown in Fig.~\ref{fig3} demonstrate
that, as for $J/U=1/4$, the pinning condition $A(0)=2/\pi$ is increasingly
violated with increasing $U$, indicating the bad-metallic nature of the phase
below the transition. Because of the smaller value of $J$ the pseudogap
is considerably narrower than for $J/U=1/4$. As a result, the characteristic
four-peaked structure in the bad-metallic phase is now better resolved
than for the ones presented in Fig.~\ref{fig2}(a). Note that, despite the
small value of $J$, the spectra shown in Fig.~\ref{fig3} differ markedly 
from the corresponding spectra of the single-band Hubbard model obtained
in the limit $J=0$. Ising Hund's exchange leads to a significant redistribution
of spectral weight at low frequencies so that even a weak coupling to the 
localized spins creates a pseudogap and destroys the zero-frequency pinning
condition. We find that these effects persist also for weak 
anisotropies $J'/J \approx 0.8\dots 0.9$~\cite{costi.07}.

In summary, we have studied  the metal insulator transition in the two-band
Hubbard model with bands of different widths and for Ising Hund's
exchange, arguing that the results for the metal insulator transition in the
wide band apply for all anisotropies $J'<J$. The present NRG-DMFT calculations 
allow a sufficiently accurate analysis of the low-frequency properties of the 
wide band in the case of anisotropic Hund's exchange. The results reveal a 
continuous quantum phase transition for
$J/U > 0.1$, i.e., in the region relevant for many transition metal oxide
materials. Close inspection of the critical properties of this transition show 
that they coincide with  the ones of the band-splitting transition
in the model with $U=0$ and variable $J$. In the region $J/U<0.1$, the transition
becomes first order with hysteresis. This can be interpreted as the 
stability of first-order transitions, in this case that at $J=0$ for small 
finite temperatures, to weak perturbations such as the 
weak coupling to localized spins. A sufficiently large perturbation
destroys this first-order behavior, giving rise, in the present case, to
a continuous quantum phase transition for $J/U>0.1$. 
The wide band is shown to be bad-metallic 
for all $J>0$, with a low-frequency pseudogap whose width is determined by $J$
and Hubbard bands whose positions are determined by $U$. The resulting four-peaked
spectral densities differ fundamentally from the familiar three-peaked structures
of the one-band model to which our model reduces in the limit $J=0$~\cite{tomita.03}. 
We have also argued that the present results for fully anisotropic Hund's exchange 
($J'=0$) are relevant to the generic case of weakly anisotropic exchange
($J'<J$) with the critical properties of the wide band transition remaining 
unaltered for $J'<J$.
\acknowledgments
We would like to thank A. Rosch for valuable discussions. Supercomputing
support from the John von Neumann Institute (J\"ulich) is gratefully acknowledged.

\end{document}